\def \SAIT #1 #2 {{\em Mem.\ Soc.\ Astron.\ It.\/} {\bf #1}, #2}
\def \MESS #1 #2 {{\em The Messenger\/} {\bf #1}, #2}
\def \ASTRNACH #1 #2 {{\em Astron. Nach.\/} {\bf #1}, #2}
\def \AAP #1 #2 {{\em Astron. Astrophys.\/} {\bf #1}, #2}
\def \AAL #1 #2 {{\em Astron. Astrophys. Lett.\/} {\bf #1}, L#2}
\def \AAR #1 #2 {{\em Astron. Astrophys. Rev.\/} {\bf #1}, #2}
\def \AAS #1 #2 {{\em Astron. Astrophys. Suppl. Ser.\/} {\bf #1}, #2}
\def \AJ #1 #2 {{\em Astron. J.\/} {\bf #1}, #2}
\def \ANNREV #1 #2 {{\em Ann. Rev. Astron. Astrophys.\/} {\bf #1}, #2}
\def \APJ #1 #2 {{\em Astrophys. J.\/} {\bf #1}, #2}
\def \APJL #1 #2 {{\em Astrophys.. J. Lett.\/} {\bf #1}, L#2}
\def \APJS #1 #2 {{\em Astrophys. J. Suppl.\/} {\bf #1}, #2}
\def \APSS #1 #2 {{\em Astrophys. Space Sci.\/} {\bf #1}, #2}
\def \ASR #1 #2 {{\em Adv. Space Res.\/} {\bf #1}, #2}
\def \BAIC #1 #2 {{\em Bull. Astron. Inst. Czechosl.\/} {\bf #1}, #2}
\def \JSQRT #1 #2 {{\em J. Quant. Spectrosc. Radiat. Transfer\/} {\bf #1}, #2}
\def \MN #1 #2 {{\em Mon. Not. R. Astr. Soc.\/} {\bf #1}, #2}
\def \MEM #1 #2 {{\em Mem. R. Astr. Soc.\/} {\bf #1}, #2}
\def \PLR #1 #2 {{\em Phys. Lett. Rev.\/} {\bf #1}, #2}
\def \PASJ #1 #2 {{\em Publ. Astron. Soc. Japan\/} {\bf #1}, #2}
\def \PASP #1 #2 {{\em Publ. Astr. Soc. Pacific\/} {\bf #1}, #2}
\def \NAT #1 #2 {{\em Nature\/} {\bf #1}, #2}

\documentstyle{memsait2002}
\input epsf.sty
\input epsfig.sty
\begin{opening}
\title{OPTICAL OBSERVATIONS OF BLAZARS: THE PERUGIA MONITORING}
\author{G. Tosti$^{1,2}$, S. Ciprini$^{1,2}$, G. Nucciarelli$^1$}
\institute{$^1$Osservatorio Astronomico, Universit$\grave{a}$ di Perugia, Italy\\
$^2$Dipartimento di Fisica,  Universit$\grave{a}$ di Perugia,
Italy}
\date{} % DO NOT INSERT ANY DATE HERE !!!
\end{opening}

\begin{document}

%\oddpagefooter{\sf Mem. S.A.It., Vol. ??, ??}{}{\thepage}
%\evenpagefooter{\thepage}{}{\sf Mem. S.A.It., Vol. ??, ??}
\oddpagefooter{}{}{} % LEAVE AS IT IS !
\evenpagefooter{}{}{} % LEAVE AS IT IS !
\
\bigskip

\begin{abstract}
In eight years, during the blazar monitoring program performed
with the Perugia automatic telescope, we were able to collect
about 20000 $BVR_cI_c$ photometric points, contributing to get
knowledge on the optical flux history of many sources brighter
than magnitude $V=17.0$. This intensive and constant optical
monitoring is essential to perform studies and multiwavelength
campains about this class of AGN. We illustrate our program and
some examples of data and preliminary results.
\end{abstract}

\section{Introduction}
Blazars are Active Galactic Nuclei (AGNs), having a variable
non-thermal continuum emission extending from radio to gamma-rays.
EGRET discovered that blazars are the most powerful extragalactic
$\gamma$-ray sources up to GeV energies. This emission is
interpreted as due to a high energy radiating plasma region
propagate downstream with shocks in an underlying jet, which
itself moves at relativistic speed and points almost directly
toward the observer (e.g. Blandford \& Rees 1978; Dondi \&
Ghisellini 1995; Kollgaard et al. 1996). Detailed studies of
blazars flux variations may provide considerable information on
the emitting region dynamics. The blazar monitoring in optical
bands permit to discriminate among the proposed emission models
and represent, together with multiwavelength observations, an
indispensable tool to investigate physical mechanism and
conditions in this objects. Moreover it's important to reach
sufficient quality, density and continuity of observations to
permit significant statistical data analysis and to establish
possible correlation with the X-ray and $\gamma$-ray emission. In
order to achieve the optical coverage necessary for this goal,
ground-bases small telescope equipped with CCD camera, can play an
important role thank to their large availability. The Automatic
Imaging Telescope (AIT) of the Perugia University is one of the
first robotic telescope in the world, and it's dedicated to blazar
observation since 1992. General information about our telescope
and blazar sample list can be found at
\verb"http://wwwospg.pg.infn.it".

\section{The Perugia Monitoring Program and Blazar Sample}

\begin{figure}[t]
\centerline{\epsfig{figure=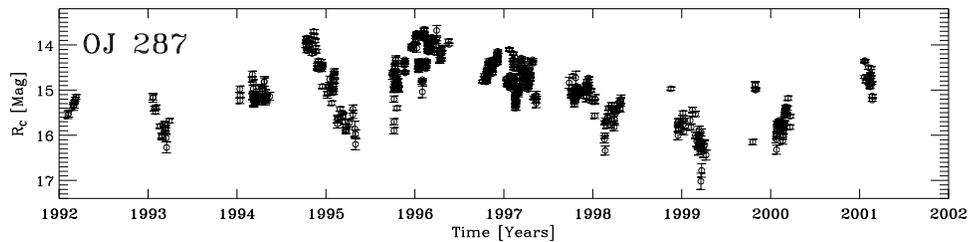,width=\linewidth,angle=90}}
\caption{\label{fig:OJ287}The light curve of OJ 287 in the $R_c$
band.}
\end{figure}

\begin{figure}
\centerline{\epsfig{figure=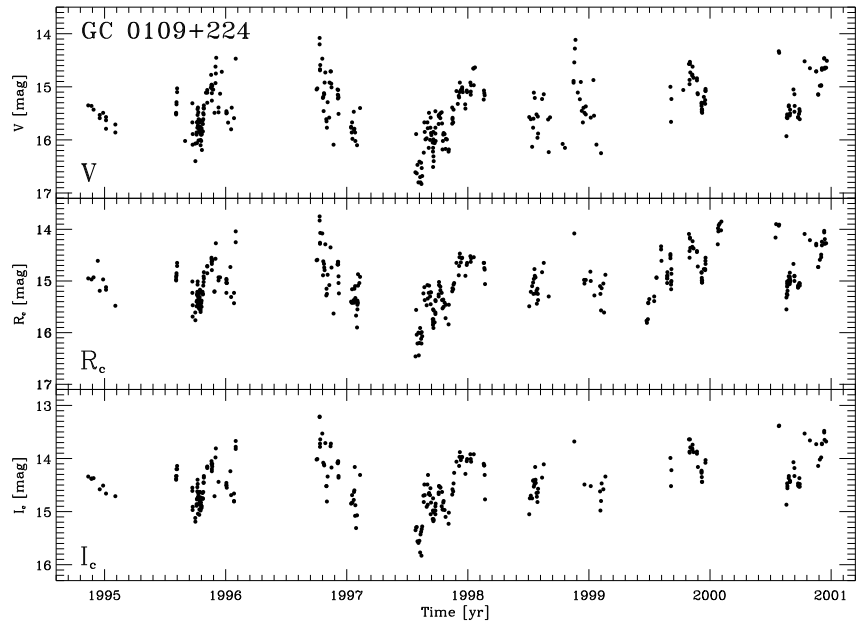,width=12cm,angle=90}}
\caption{\label{fig:0109}Example of a well monitored source: GC
0109+224.}
\end{figure}

The AIT is based on an equatorially mounted 0.4-m Newtonian
reflector having a 0.15-m refractor. A CCD camera and
Johnson-Cousins $BVR_cI_c$ filters are utilized for photometry.
The system, become full operative in October 1994. It is equipped
with an autoguider (based on a CCD camera) that is mounted at the
focus of the refractor. The autoguider is also used to verify the
correctness of each pointing and to monitor sky conditions.
\begin{figure}
\centerline{\epsfig{figure=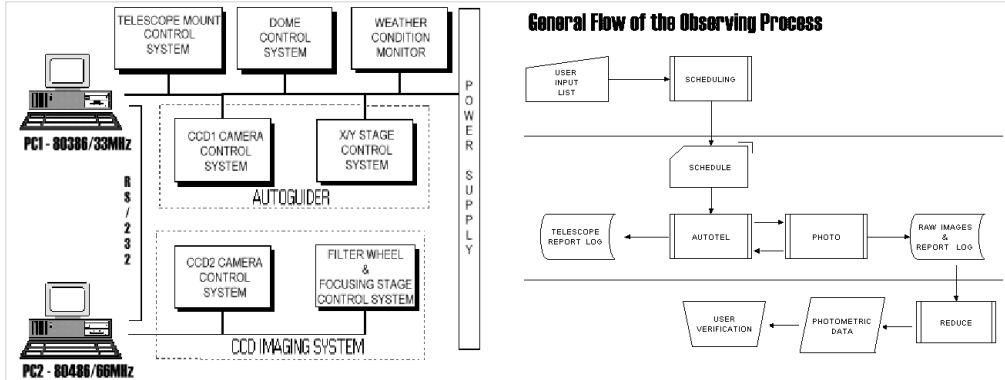,width=\linewidth,angle=0}}
\caption{\label{fig:system}Functional block diagram of the AIT
system and the observing process.}
\end{figure}
The control of the observatory during the night is now completely
automated with the computers deciding the opening or the closing
of the dome shutter, selecting the object from the observing list,
setting and centering the telescope on the field, taking the
desired CCD exposures, and reducing the data after the end of the
night (Tosti, Pascolini \& Fiorucci 1996).
\begin{figure}[b]
\centerline{\epsfig{figure=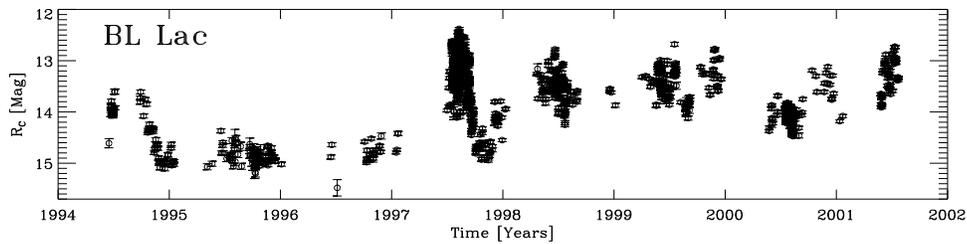,width=\linewidth,angle=90}}
\caption{\label{fig:BLLac}The light curve of BL Lac in the $R_c$
band.}
\end{figure}
Our blazar sample contain about 40 blazar, contributing to get
knowledge on the variability of many these sources. In Fig.
\ref{fig:BLLac} and \ref{fig:OJ287} we report two example of light
curve in $R_c$ band, while in Fig. \ref{fig:0109} one example of a
less famous source but indeed well monitored by us in three band.
\section{Results and conclusions}
The relevant amount of data from the best monitored sources, allow
us to perform a first statistical analysis to investigate the time
variability characteristics and the light curves. We used the
Structure Function (SF) defined as $SF_{\Delta t}=\frac{1}{N}\sum
\bigl(F(t)-F(t+\Delta t)\bigl)^2$, where $\Delta t$ is the time
lag and $F$ the flux, in a $\log(\Delta t)$ versus $\log(SF)$
plot. It's similar to a sort of power spectrum suitable for
nonperiodic data (see, e. g., Hufnagel \& Bregman 1992).
%%%%%%%%%%%%%%%%%%%%%%%%%%%%%%%%%%%
The SF slope $\alpha$ is an interesting parameter because it
highlights some basic properties of the light curves (variability
like shot noise, like flickering noise etc...). In our sample the
optical variability was generally characterized by an intermediate
state between shot and flickering noise ($\alpha=0.7\pm2$). Shot
noise has an infinite memory of all the preceding events
($\alpha=1$), on the other hand flickering noise depends
especially from the last events ($\alpha\rightarrow0$) (Fiorucci
et al. 1999). Our result may suggest that the optical flares are
produced in small-scale structures within the jet, such us shocks
or plasma turbulence.
\par
To optimize the blazar optical monitoring, important for
developing a statistical data analysis and for correlation
multiwavelength analysis, we need a stable collaboration among
observers, an automatization of the telescopes and data reduction,
a standard photometric and database systems (e.g. Villata \&
Raiteri 2001, Mattox 1999).

% References. We avoided using the \bibitem commmand since we found it is
% somewhat platform-dependent. We also avoided using the \cite{keyword}
% command since we found it cumbersome. However, if you are an expert
% LateX user you may use the various LateX tools for the references
% provided they give the same printout formats of the examples given here.

\vskip 0.8cm

\noindent {\bf DISCUSSION}

\vskip 0.4cm \noindent {\bf J. BECKMAN:} I am very impressed with
the quality and quantity of the observation made with a relatively
small telecope. There must be many potential observers who could
collaborate in such programm. What would you estimate to be the
smallest telescope which could usefully take part in this kind of
monitoring?

\vskip 0.3cm \noindent {\bf G. TOSTI:} The smallest configuration
useful to the blazar optical monitoring is a telescope with
diameter of 25-30 cm equipped with a CCD camera and only a $R_c$
(Cousins) filter.

\vskip 0.3cm \noindent {\bf N. SMITH:} How frequently do you
observe ``single'' data points which are significantly above the
surrounding data points and how real do you believe them to be? Do
you see such data points in the lightcurves of your comparision
stars?

\vskip 0.3cm \noindent {\bf G. TOSTI:} Sparse photometric points
depend on the sampling condition (weather and other) but logs of
observations and the CCD images are controlled one by one.
Moreover we can compare them with the observation in four-three
filter and relevant peaks are confronted with data of other
observatories.

\end{document}